\documentclass{article}
\usepackage{epsf}
\usepackage{amssymb}
\usepackage[T2A]{fontenc}
\usepackage[english]{babel}
\usepackage[dvips]{graphicx}

\begin{document}
\baselineskip5mm

\title{Quasi-Mandelbrot sets for perturbed complex analytic maps: visual patterns}

\author{A.V.Toporensky}
\date{}
\maketitle
\hspace{8mm} {\em Sternberg Astronomical Institute, Moscow University, Moscow 119992, Russia}

\begin{abstract}
We consider perturbations of the complex quadratic map $ z \to z^2 +c$ and
corresponding changes in their quasi-Mandelbrot sets. Depending on particular 
perturbation, visual forms of quasi-Mandelbrot set changes either sharply
(when the perturbation reaches some critical value) or continuously. In the latter
case we have a smooth transition from the classical form of the set to some forms, constructed
from mostly linear structures, as it is typical for two-dimensional real number dynamics.
Two examples of continuous evolution of the quasi-Mandelbrot set are described.
\end{abstract}

Complex dynamics is a well-known source of beautiful images which 
bring this branch of pure mathematics into attention of wide groups
of non-mathematicians, and provide interesting connections between math 
and art (see, for example \cite{BF} and references therein).
On the other hand, the situation with real two-dimensional dynamics is
more complicated. There are many examples of artistic interest (\cite{BF, Sprott, bs}),
however, they may be considered as an "island" in a wide "sea" of other systems
without any esthetic value. 

It is also remarkable that visual patterns given by complex and real dynamics are
quite different. Mostly based on round structures, typical Julia sets of
complex maps are sometimes compared with "baroque architecture", while analogs
of Julia sets and basins of attraction in the case of real maps usually consist of
quasi-parallel bands, long tongues, and other linear structures. 

However, from the mathematical viewpoint one-dimensional complex- and two-dimensional
real maps are not completely different entities: each 1-dim complex map can be
considered as 2-dim real map by, for example, separating real and imaginary parts.
For example, the simplest nonlinear complex map (and the most famous one)
\begin{equation}
z \to z^2 +c
\end{equation}
is equivalent to a real map
\begin{equation}
x \to x^2 - y^2 + c_1,
\end{equation}
\begin{equation}
y \to 2 xy + c_2,
\end{equation}
where $c_1$ and $c_2$ are the real and imaginary parts of complex parameter $c$.

In general complex analytic maps form a particular subset of two-dimensional real maps which satisfy
the Cauchy - Riemann conditions. The goal of our study is to understand what 
happens in a zone located 
in some sense "near" this subset.
We consider one-parameter family of maps
having a complex analytic structure for a particular value of the parameter, and
then study properties of the map when the parameter is close, but is not
equal to this particular value. Pure mathematical aspects of this problem 
have been studied in a number of papers \cite{b,Peckham1,Peckham2}, 
though a general picture is still far from
being well understood. In the present paper we address more informal question of
visual appearance of sets created by such maps, and try to see
a transition from pure complex visual forms to those typical to a real
dynamics.
 
Our starting point is the quadratic map (1). Instead of fixing $c$ to consider a sequence
of corresponding Julia sets  we have chosen the Mandelbrot set as the pattern to perturb.
There are two reason for this. One is technical simplicity of this set to calculate,
the second is the wide-spread popularity of the Mandelbrot set and its role in bringing
mathematical achievements to a more general public.

There two definition of the Mandelbrot set which are equivalent 
for complex analytic maps.
First, in is the set of $c$ for which the Julia set of the map (1) is connected.
Second, it is the set of $c$ for which the orbit of the zero-point of the map (1) does 
not escape to infinity. We take the second definition to calculate the (quasi)-Mandelbrot set
for real maps, using the name "connectedness locus" 
for the set created according to the first definition. We leave studies of the connectedness loci
as well as Julia sets for real perturbations of complex maps for a future work.

To start our investigations we should fix the perturbations imposed
on the quadratic map. Some maps do not show continuous transitions to visual patterns of
real dynamics, as
it can be seen by close look on the map
\begin{equation}
x \to x^2 -\epsilon(y-x)^2 +\lambda_1
\end{equation}
\begin{equation}
y \to y^2 -\epsilon(x-y)^2 +\lambda_2
\end{equation}
studied in \cite{Saratov}. For $\epsilon=0.5$ this system is equivalent to the map (1)
with the following redefinition of variables: $x=Re(z)+Im(z)$, $y=Re(z)-Im(z)$, $\lambda_1=
Re(c)+Im(c)$, $\lambda_2=Re(c)-Im(c)$.
In order to understand what happens for other values of the parameter 
$\epsilon$ we remind a reader some 
simple facts from general theory of two-dimensional real algebras  (see,
for example, \cite{Yaglom}).
To define a particular algebra of this type it is enough to fix the square of
imaginary unit: $ i^2 = a + bi$, where $a$ and $b$ are real numbers. There are
three qualitatively different possibilities:
\begin{itemize}
\item If $p + q^2/4 =-k^2<0$ the resulting algebra is isomorphic to the algebra of complex
numbers with the imaginary unit $J$, and the isomorphism has the form 
$i=q/2+kJ$.
\item If $p+q^2/4=0$ the resulting algebra is isomorphic to the algebra of dual numbers
with the condition $J^2=0$.
\item If $p>q^2/4$ the resulting algebra is isomorphic to the algebra of perplex numbers
with the condition $J^2=1$.
\end {itemize}
Sometimes the perplex numbers are called as double numbers, split-complex numbers or
hyperbolic numbers (in the latter case two other number systems are correspondingly 
elliptic and parabolic numbers).

The map (4-5) is specially constructed so that it coincides
with the initial quadratic map (1) for perturbed two-dimensional algebra. If $\epsilon > 0.25$
this algebra is isomorphic to the algebra of complex numbers, and the Mandelbrot set for the map (4-5)
is nothing else but the "classical" Mandelbrot set transformed according to 
corresponding linear transformation.
For $\epsilon=0.25$ an abrupt transition occurs, and we obtain and infinite 
band  -- an analog of the Mandelbrot
set for dual numbers. For smaller $c$ the result is a rectangular 
representing this set
for the algebra of perplex numbers. 
The abrupt nature of transition for $\epsilon=0.25$ makes the map (4-5) useless for our present study.      

The map (4-5) represents a rather peculiar example, and the other known 
map, studied by Peckhman in \cite{Peckham1, Peckham2} gives us a success. This map conserves
the meaning of $z$ as a complex variable with real part $x$ and imaginary part $y$, however,
it violates the Cauchy-Riemann conditions by adding a linear term depending on $ z $-conjugate:
\begin{equation}
z \to z^2 +c + a \bar z,
\end{equation}
or, in real variables
\begin{equation}
x \to x^2 - y^2 + c_1 + ax
\end{equation}
\begin{equation}
y \to 2xy + c_2 - ay.
\end {equation}

In Figs. 1 -- 6 we present several pictures of a part of quasi-Mandelbrot set boundary in the lower - right
quadrant for $a$ from $0.1$ to $0.5$ with the step equal to $0.1$. We can see clearly how the
deviation from the famous structure of the Mandelbrot set, touching the set 
at it outer range for $a=0.1$, then goes deep inside and transform
it into the structure typical for real maps. At $a=0.5$ almost all 
structure remaining from complex origin of the map under investigation 
is washed out.

Qualitatively similar patterns appear for negative $a$ in the same range of $|a|$, so we do not
present corresponding pictures here.


Figure 1: A part of the "original" Mandelbrot set boundary with no perturbation.


Figure 2: A fragment of quasi-Mandelbrot set boundary with $a=0.1$. The distortion have touched 
an outer part of a bulb.



Figure 3: The same boundary for $a=0.2$. Linear structures have been formed
in the outer part of the bulb.


Figure 4: The same boundary for $a=0.3$. The bulb have been almost completely
destroyed.


Figure 5: The same boundary for $a=0.4$. Only remnants of the non-perturbed
structures are visible.


Figure 6: The same boundary for $a=0.5$. Almost all complex analytic
structure have been washed out.

Finally, we consider a family of maps which, to our knowledge, is not studied previously.
It is possibly the simplest way to disturb the map (2-3), 
though from the viewpoint of complex dynamics it looks crazy. This map has the form
\begin{equation}
x \to x^2 - y^2 + c_1
\end{equation}
\begin{equation}
y \to \alpha x y + c_2
\end{equation}
and represents a complex analytic map for $\alpha=2$. The first equation
of the real decomposition of the complex quadratic map (2) remains unchanged.

We start with an auxiliary map which has the same equation (9) for the
$x$-component and represent the quadratic map for a 
disturbed algebra with $i^2=-1+bi$. It has the 
following equation for $y$
\begin{equation}
y \to 2 x y + b y^2  + c_2.
\end{equation}

The properties of the map (9,11) is similar to those for (4,5), and the
algebra becomes isomorphic to perplex numbers for $b<-2$ showing a sharp
change of its quasi-Mandelbrot set for $b=-2$.
On the other side, positive $b$ keep the 
considered algebra within the class of complex numbers. 
The map (9,10) is not equivalent to (9,11), 
though, as the fate of a particular trajectory is typically
determined by its behavior at the range of variables $x$ and $y$ of the order of unity,we may
expect that  it should be a remarkable change of dynamics for $\alpha$ being somewhere in the interval
$(0, 2)$.
Our numerical experiment confirms these considerations. We find qualitative changes
in the shape of the quasi-Mandelbrot set for $\alpha$ less than unity. In Figs. 7 -- 10 some of "snapshots"
of the boundary are presented. At the final
picture created for $\alpha=0.5$ we can see a typical pattern of 
a real number dynamics. 
It is also remarkable
that unlike previous case of the map (6), the real-like features of the map(9,10) 
do not propagate
from the outer part of the set, 
but steadily develop in the whole range of the boundary structure,
which becomes more and more elongated with decreasing $\alpha$.

On the other hand, for $\alpha>2$ the change is rather slow
and not so interesting from the viewpoint of visual impression. Even
for $\alpha=10$ some structure remaining from the original Mandelbrot set still survives,
though linear structures are well developed for this value of $\alpha$.


Figure 7: A part of quasi-Mandelbrot set boundary for the map (9) -- (10) with
$\alpha=1.1$. The whole structure is slightly disturbed.


Figure 8: The same boundary for $\alpha=0.9$. Bulbs have been deformed, but
not destroyed.


Figure 9: The same boundary for $\alpha=0.7$. Bulbs have been deformed into
quasi-linear structures.


Figure 10: The same boundary for $\alpha=0.5$. Bulbs became elements of
linear structures.

In the present paper we have studied visual appearance of quasi-Mandelbrot
sets for perturbed quadratic complex map. Two different cases can be
distinguished. In a particular case  when
a perturbed map is equivalent to the initial one for a perturbed
two-dimensional real algebra, the form of the quasi-Mandelbrot set change
sharply reflecting change in the isomorphic class of the algebra.
For general perturbations, however, we expect a smooth transition from
complex to real visual patterns. Two studied in the present paper 
maps show this transition clearly.

Our present work can be pushed further in several ways.
Continuous nature of transition can be used in constructing "movies" showing how one fractal
structure is replaced by another, which may make the beauty of fractals dynamical. 
Another possible application of
the transition range of parameters found in the present paper is creation
of generalized Julia sets, which can provide interesting changes in their visual patterns.
Other possible directions include non-polynomial perturbations of complex analytical
maps which have already been studied in mathematical literature \cite{b}.   

\section*{Acknowledgments}
Author is grateful to Julien Clinton Sprott for discussions.

\end{document}